\documentclass{mn2e}
\usepackage{amssymb}
\usepackage{ulem}
\usepackage{color}
\input epsf.sty
\newif\ifAMStwofonts
\AMStwofontstrue

\newcommand{\kms}{km s$^{-1}$}
\def\gax    {${_>\atop^{\sim}}$}

\begin{document}

\title[WLQ SDSS J094533.99+100950.1]
{SDSS J094533.99+100950.1 - the remarkable weak emission line quasar} 

\author[Hryniewicz et al.]
  {K.~Hryniewicz$^{1,2}$, B.~Czerny$^1$, M. Niko\l ajuk$^2$ and
  J. Kuraszkiewicz$^3$\\
  $^1$Nicolaus Copernicus Astronomical Center, Bartycka 18, 
	00-716 Warsaw, Poland\\ 
  $^2$Faculty of Physics, University of Bia\l ystok, Lipowa 41, 
	15-424 Bia\l ystok, Poland \\ 
  $^3$Harvard-Smithsonian Center for Astrophysics, 60 Garden Street,
	Cambridge, MA 02138, USA}

\maketitle
\begin{abstract}

Weak emission line quasars are a rare and puzzling group of
objects. In this paper we present one more object of this class found
in the Sloan Digital Sky Survey (SDSS). The quasar SDSS
J094533.99+100950.1, lying at $z = 1.66$, has practically no C\,IV
emission line, a red continuum very similar to the second steepest of
the quasar composite spectra of Richards et al., is not strongly
affected by absorption and the Mg\,II line, although relatively weak,
is strong enough to measure the black hole mass. The Eddington ratio
in this object is about 0.45, and the line properties are not
consistent with the trends expected at high accretion rates. We
propose that the most probable explanation of the line properties in
this object, and perhaps in all weak emission line quasars, is that
the quasar activity has just started. A disk wind is freshly launched
so the low ionization lines which form close to the disk surface are
already observed but the wind has not yet reached the regions where
high ionization lines or narrow line components are formed. The
relatively high occurrence of such a phenomenon may additionally
indicate that the quasar active phase consists of several sub-phases,
each starting with a fresh build-up of the Broad Line Region.
\end{abstract}



\section{Introduction}

Strong, broad emission lines are the characteristic signature of
Active Galactic Nuclei (AGN). However, the Sloan Digital Sky Survey
(SDSS) finds, among thousands of new AGN, rare objects with very weak
or almost absent emission lines (Fan et al. 1999, Shemmer et al. 2009,
Diamond-Stanic et al. 2009, Plotkin et al. 2010). Most of these
objects, forming a new class of weak line quasars (hereafter referred
to as WLQs), have been found primary at high redshifts ($z$\gax 2).
A few sources lying relatively close by have been studied in greater
detail: PHL 1811 ($z = 0.19$; Leighly et al. 2007), PG 1407+265 ($z =
0.94$; McDowell et al. 1995), and HE 0141-3932 (z=1.80; Reimers et
al. 2005).  The study of WLQs optical/UV continuum properties, X-ray
and radio emission shows no obvious difference from typical emission
line quasars and no apparent similarity to the weak-line BL Lac
objects (Shemmer et al. 2009).

No generally accepted explanation for the weakness/absence of emission
lines has been found so far. Possibilities so far considered include:
1) dust obscuration or extreme broad absorption line (BAL) effect,
which is unlikely since there are no signs of the deep and broad
absorption troughs in the spectra (Anderson et al. 2001, Collinge et
al. 2005) and the X-ray absorbing column is low compared with BAL QSOs
($N_{\rm H} < 5 \times 10^{22}$ cm$^{-2}$, Shemmer et al. 2009). 2)
Gravitational lensing does not remove the emission lines (Shemmer et
al. 2006), and strong C\,III] and C\,IV lines are found in
gravitationally amplified quasars (e.g. Dobrzycki et al. 1999, Wayth
et al. 2005). 3) Relativistic beaming is a good explanation for the
absence of strong lines in BL Lac objects but WLQs, in contrast to BL
Lacs, have bluer optical/UV continua, are radio quiet and show no
variability or strong polarization (Shemmer et al. 2009,
Diamond-Stanic et al. 2009).  4) The difference in the ionizing
continuum also faces difficulties since at least the observed part of
the continuum in those sources looks typical for a quasar. However,
those objects may still emit more vigorously in the unobserved UV,
thus having a higher accretion rate than the average quasars, which
may affect the lines, as e.g. seen through the Baldwin effect. No
black hole mass determination or Eddington ratio was possible for the
sample of objects analyzed by Shemmer et al. (2009) so as to confirm
or reject this hypothesis. In this paper we present a new WLQ object
found in the SDSS survey which opens a possibility to address this
issue.

\section{Selection procedures and observed properties of 
the quasar SDSS J094533.99+100950.1}

\begin{figure*}
\epsfysize=125mm
\epsfbox{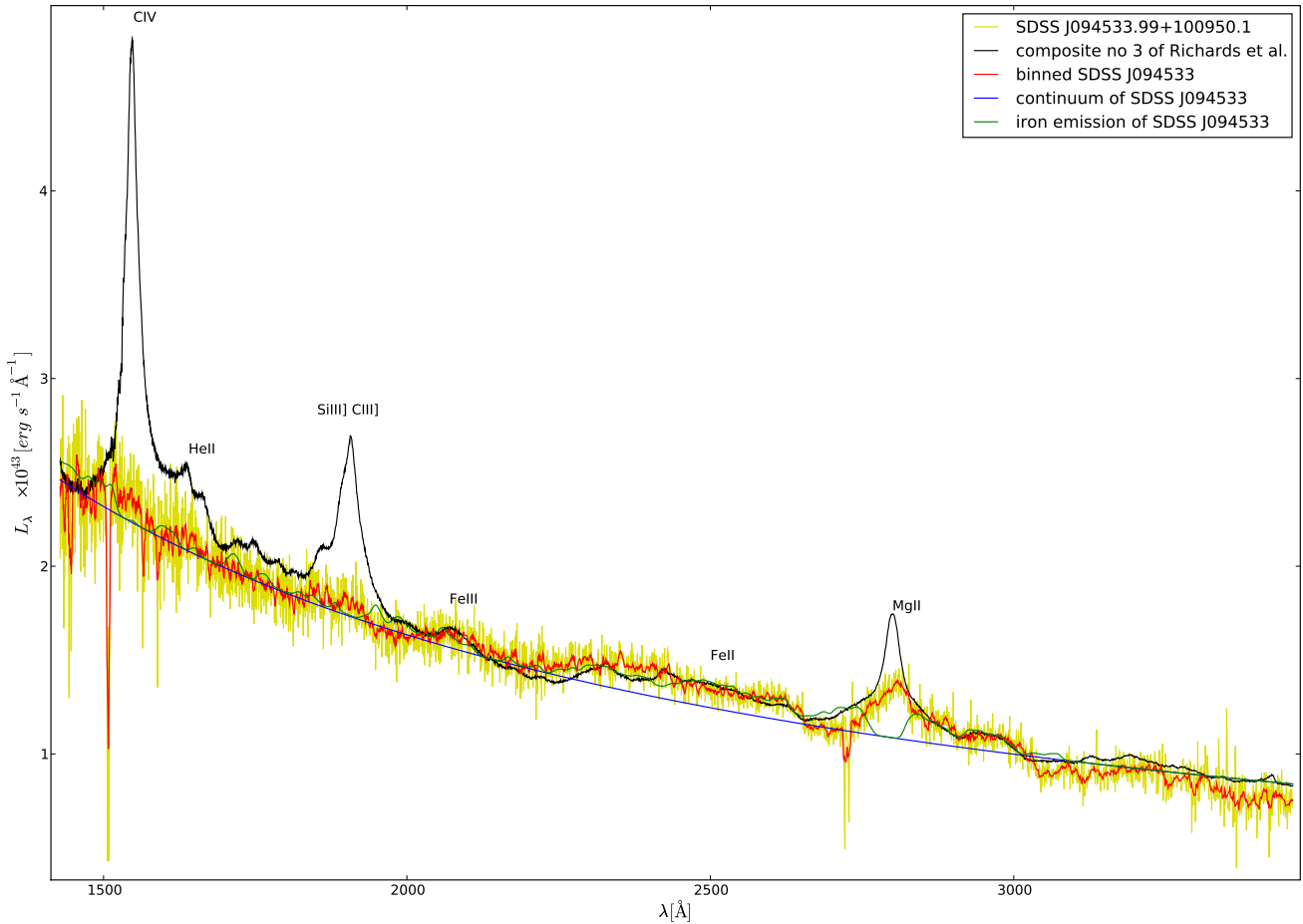}
\caption{The restframe spectrum of SDSS J094533.99+100950.1 corrected
for Galactic reddening (yellow line), the binned spectrum of the
quasar (red line), its fitted, underlying continuum (blue line) and
iron emission (green line). For comparison, in black, the Richards et
al. (2003) composite spectrum (no.~3) is shown, greyshifted to match
the SDSS J094533.99+100950.1 spectrum.}
\label{fig:spec}
\end{figure*}

SDSS J094533.99+100950.1 was serendipitously discovered by us while
searching the full SDSS DR5 (Adelman-McCarthy et al. 2007) spectral
database with the aim of finding interesting quasars for which
atypical line and/or continuum properties would prohibit them being
found and classified as quasars by the SDSS automatic pipeline. In
particular, we looked for quasars with extreme blue continua (these
would have been missed due to colour cuts used by the SDSS to avoid
regions of colour space dominated by white dwarfs, A-stars and M
star+white dwarf pairs - see Richards et al. 2002) and quasars with
normal (unreddened by dust) continua but with non-standard emission
line properties. We analyzed all, 1\,048\,960, spectra available in
the SDSS DR5 spectral database, classified by the SDSS automatic
procedures into: galaxies, quasars, stars, sky, and unclassifiable
objects (see Adelman-McCarthy et al. 2007).


The spectrum of SDSS~J094533.99+100950.1 appeared as particularly
interesting, showing a normal quasar continuum but with atypical, weak
emission lines, with practically nonexistent C\,IV and C\,III], and
weak, although clearly visible, Mg\,II.  The object was identified as
a quasar by the SDSS pipeline due to the presence of the Mg\,II line,
and is hence also included in the Schneider et al. (2007) SDSS quasar
catalogue.  The spatial coordinates are $\alpha = 146.39163$, $\delta
= 10.163927$ and the determined redshift is $z=1.66160\pm 0.00025$.
The quasar's spectrum was observed on April 25, 2003 and classified as
``excellent'' by the SDSS database.

\subsection{Emission line measurements - modelling of the spectrum}

In order to study the weak emission line properties of our WLQ we
modeled the spectrum by fitting a power law to the underlying
continuum, accounting for blended iron emission (the iron template of
Vestergaard \& Wilkes 2001 was used) and fitting single Gaussians to
the emission and absorption lines.  Atmospheric emission lines that
still remained after the automatic pipeline reduction were first
removed, and the spectrum was corrected for Galactic reddening using
the Cardelli et al. (1989) extinction curve, adopting a colour excess
of $E(B-V) = 0.06$ determined from the Galactic neutral hydrogen
column from Dickey \& Lockman (1990) and Stark et al. (1992), and
assuming a fixed conversion of $N(HI)/E(B-V) = 5.0 \times 10^{21}
$cm$^{-2}$ mag$^{-1}$ (Burstein \& Heiles 1978).

\begin{table}
\caption{Continuum and iron fitting windows
}
\begin{tabular}{cc}
\hline 
\hline
Model & Fitting windows  \\
Name & rest frame wavelength range in \AA  \\

\hline
Cont$^1$ & 1455-1470, 1690-1700, 2160-2180, 2225-2250,  \\
      & 3010-3040, 3240-3270  \\

\hline
Fe1$^2$ & 2020-2120, 2250-2340, 2440-2650, 2900-3000 \\

\hline
Fe2$^3$ & 2020-2120, 2250-2300, 2500-2650, 2850-3000 \\

\hline
Fe3$^4$ & 1490-1502, 1705-1730, 1760-1800, 2250-2320 \\
      & 2470-2625, 2675-2715, 2735-2755, 2855-3010  \\

\hline
ContFe4$^5$ & 1455-1470, 1685-1700, 2020-2120, 2190-2210 \\
        & 2250-2300, 2450-2650, 2850-3070  \\
\hline
ContFe5$^6$ & 1440-1470, 1700-1820, 1950-2400, 2450-2700 \\
        & 2850-3100  \\ 
\hline 
ContFe6$^7$ & 1455-1470, 1690-1700, 2020-2120, 2160-2180 \\
   	 & 2225-2650, 2900-3000, 3010-3040, 3240-3270 \\
	 & 3350-3400 \\

\hline
\end{tabular}
\\
\begin{flushleft}
Column (1) - name of model; column (2) - wavelength range of spectral
windows in which the model was fit.  

$^1$ - continuum fitted as a power law in Forster et al. (2001)
continuum windows;  \\ 
$^2$ - iron template fitted in Forster et al. (2001) iron spectral
windows, excluding wavelength range 2340-2440\AA\,
(atmosphere artifact + feature at 2350\AA); \\
$^3$ - iron template fitted in Forster et al. (2001) iron 
windows, with regions around C\,II]\,$\lambda2327$ and
O\,II\,$\lambda2441$ excluded, and 2850-2900\AA\ window 
(at red side of Mg\,II) added for a more consistent fit; \\
$^4$ - iron template fitted in spectral windows defined by Vestergaard \&
Wilkes (2001), with regions of C\,II] emission and  narrow
absorption at  C\,IV (1427-1505\AA) and Mg\,II (2715-2735\AA) excluded; \\
$^5$ - power law continuum and iron emission fitted simultaneously
in the  ``Cont''  and  ``Fe1''  windows (some windows are narrower or
slightly shifted to better match the model with the spectrum); \\
$^6$ - power law continuum and iron emission fitted
simultaneously in optimum windows chosen by us and created by
comparison of the Richards et al. (2003) composites and Vestergaard \&
Wilkes (2001) iron template;  \\
$^7$ - power law continuum and iron emission fitted
simultaneously in the ``Cont'' model windows complemented with a
3350-3400\AA\ window and iron windows from Forster et al. (2001) .

\end{flushleft}
\label{tab:OurWindows}
\end{table}


The underlying power law continuum was fitted to regions of spectrum
uncontaminated by emission lines and away from blended iron
emission. The continuum windows used are shown in
Table\ref{tab:OurWindows} (model ``Cont'') and taken from Forster et
al. (2001). Blended iron emission was modeled using the UV iron
template of Vestergaard \& Wilkes (2001) to spectral windows defined
in Forster et al. (2001) (Table\ref{tab:OurWindows} model ``Fe1''). We
also experimented with slightly modified iron windows which excluded
spectral regions of C\,II]\,$\lambda2327$ and O\,II\,$\lambda$2441
emission and/or C\,IV and Mg\,II NAL absorption (models ``Fe2'' and
``Fe3'' in Table\ref{tab:OurWindows}). We present the results of our
continuum and iron modelling in Table~\ref{tab:continuum}.

In model no. 1 (Cont-Fe1-best in Table~\ref{tab:continuum}) we
followed the procedure described in Vestergaard \& Wilkes (2001), where
first the underlying power law continuum was fitted and subtracted
from the spectrum, followed by the modelling of iron emission.  The
Vestergaard \& Wilkes (2001) iron template spectrum was broadened by
convolving with Gaussian functions of widths between 900~km\,s$^{-1}$
and 9000~km\,s$^{-1}$ and separated by steps of 250~km\,s$^{-1}$,
while conserving the total flux in each template.  We chose the
broadening value for which $\chi^2$ was lowest.  Finally, we
subtracted the best iron fit from the spectrum and again fitted the
continuum. This procedure was repeated two more times.
Final parameters are presented in Table~\ref{tab:continuum}.

In models no. 2-4 (Cont-Fe1,2,3) a similar procedure was used for the
continuum fitting, however broadening of the iron template and the
scaling factor of the amplitude was estimated automatically. We
experimented with standard and slightly modified iron windows (models
Fe1, Fe2, Fe3 from Table\ref{tab:OurWindows}).

In models no. 5-7 (i.e. ContFe4,5,6) power law continuum and iron
template were fitted simultaneously (as a sum) to spectral windows
including standard and slightly modified continuum and iron windows
from Forster et al. (2001) (models ``ContFe4'', ''ContFe5'' and
``ContFe6'' in Table~\ref{tab:OurWindows}).

\begin{figure}
\epsfxsize=8 cm
\epsfbox{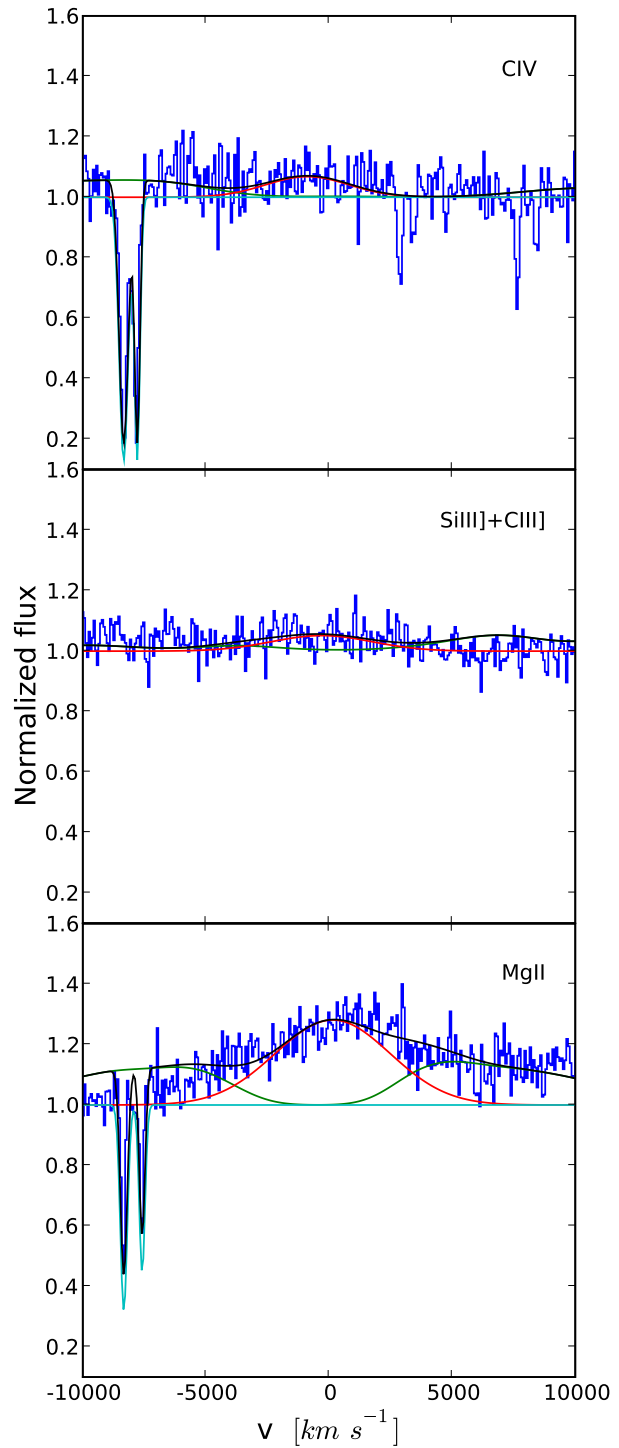}
\caption{The regions of the C\,IV, Si\,III]+C\,III] and Mg\,II lines in 
the spectrum of SDSS J094533.99+100950.1 (blue).
Green line shows the broadened (FWHM(Fe\,II)=1650 \kms) iron emission
template, red and cyan show fitted emission and absorption lines
respectively, while the black line is sum of all components. Zero
velocity was at the theoretical wavelength of each line at a given
redshift. The plotted flux was normalized by the continuum.}
\label{fig:lines}
\end{figure}

The best continuum and iron emission fit, is represented by model
no. 1 (Cont-Fe1-best). It has the lowest reduced $\chi^2$ value and
visually gives the best fit to iron emission in the vicinity of the
Mg\,II line, which then leads to the best Mg\,II emission line fit (as
shown by the lowest $\chi^2$ and weakest residuals).  In
Fig.~\ref{fig:spec} we plot this continuum and iron emission fit over
the restframe, Galaxy dereddened spectrum of
SDSS~J094533.99+100950.1. The fit is also included in
Fig.~\ref{fig:lines} and the emission line and central engine
properties derived further in the text are made using this model.

The best fit continuum has a power law index, $\alpha_\lambda$
($f_{\lambda} \propto \nu^{\alpha_\lambda}$) equal to $-1.232 \pm
0.017$, which is equivalent to $\alpha_{\nu} = -0.77$ ($f_{\nu} \propto
\nu^{\alpha_\nu}$).
Therefore, the SDSS J094533.99+100950.1 continuum is, in this respect,
almost identical to the no.~3 quasar composite from Richards et
al. (2003). However, we also compared quantitatively the underlying
continua of SDSS J094533.99+100950.1 and Richards et al. (2003)
composites no.~1-4 by fitting the latter with a power law (to the same
spectral windows as for our WLQ - the ``Cont'' model from Table~1) and
allowing for internal reddening.
We checked both Gaskell et al. (2004) and SMC-bar (Sofia et
al. 2006) extinction curves.  We modified the composites and allowed
for either reddening or dereddening.  The lowest $\chi^2$ was obtained
for composite 4 dereddened by $E(B-V)=0.75$ for Gaskell et al. (2004)
extinction, which is relatively high due to its weak dependence on
wavelength, which thus requires large reddening to produce any
significant change in spectral shape. For the case of SMC-bar
extinction curve a lower value is needed, $E(B-V)=0.02$.  For
composite 3 of Richards et al. (2003) no reddening/dereddening was
required an acceptable fit. For the other two composites reddening was
required but the fit was always worse than for pure, unreddened
composite 3. Since the exact wavelength dependence of the intrinsic
reddening in quasars is still under debate, we decided to use
composite 3 for comparison.  In Fig.~\ref{fig:spec} we show the
Richards et al. (2003) composite no.~3. plotted over the rest frame
SDSS J094533.99+100950.1 spectrum, corrected for Galactic extinction.

Despite the almost identical continua, the two spectra 
are widely different with respect to the emission line properties.
The Mg\,II emission line is clearly present but the narrow component
in the line profile is absent. The contribution of the Fe\,II and
Fe\,III is almost typical, while the Si\,III]+C\,III] and C\,IV
emission is extremely weak compared with a typical quasar.
There are also two deep but narrow absorption components (clearly
resolved into doublets in the unbinned data plot), blueshifted by
$\sim 8300$ \kms with respect to the C\,IV and Mg\,II emission line
position expected from WLQ's adopted redshift.

\begin{table*}
\caption{Properties of fitted continuum and iron emission template for
different fitting windows.
}
\begin{tabular}{clccccccc}
\hline 
\hline
No & &\multicolumn{2}{c}{\underline{\hspace{6em}continuum\hspace{6em}}} 
 & \multicolumn{3}{c}{\underline{\hspace{7em}Fe\,II
template\hspace{7em}}} & $\chi^2$/ndf. & fitting \\
 & & $\alpha_{\lambda}$ & F$_{3000\AA}$  & FWHM & scale factor & REW & & windows \\
 & & & [erg s$^{-1}$ cm$^{-2}$ \AA $^{-1}$] & [km s$^{-1}$] & & [\AA] & &\\
\hline

1.& \multicolumn{2}{l}{\bf Cont-Fe1-best}& & & & & & \\ 
  & Cont: & $-1.232 \pm 0.017$ & $(5.24 \pm 0.69)\cdot 10^{-16}$ & --- & --- & --- & 1.51 & Cont\\
  & Fe: &  --- & --- & $1650 \pm 130$ & $0.01176 \pm 0.00027$ & $82$ &
1.30 & Fe1\\

\hline

2.&\multicolumn{2}{l}{Cont-Fe1}& & & & & &\\
  & Cont:& $-1.233 \pm 0.017$ & $(5.24 \pm 0.69)\cdot 10^{-16}$ & --- & --- & --- & 1.56 & Cont\\
  & Fe: &  --- & --- & $2150 \pm 380$ & $0.01139 \pm 0.00027$ & $79$ &
1.30 & Fe1\\ 

\hline
3.&\multicolumn{2}{l}{Cont-Fe2}& & & & & &\\ 
  & Cont:& $-1.233 \pm 0.017$ & $(5.24 \pm 0.69)\cdot 10^{-16}$ & --- & --- & --- & 1.58 & Cont\\
  & Fe: &  --- & --- & $2680 \pm 420$ & $0.01208 \pm 0.00027$ & $84$ & 
1.30 & Fe2\\

\hline
4.&\multicolumn{2}{l}{Cont-Fe3}& & & & & &\\
  & Cont: & $-1.240 \pm 0.017$ & $(5.24 \pm 0.69)\cdot 10^{-16}$ & --- & --- & --- & 1.58 & Cont\\
  & Fe: & --- & --- & $3440 \pm 550$ & $0.01070 \pm 0.00027$ & 75 & 
1.57 & Fe3 \\

\hline
5.&\multicolumn{2}{l}{ContFe4}& & & & & &\\ 
& Cont+Fe: & $-1.194 \pm 0.011$ & $(5.40 \pm 0.43)\cdot 10^{-16}$ & 
$1140 \pm 190$ & $0.00849 \pm 0.00043$ & $58 $ & 1.55 & ContFe4\\

\hline
6.&\multicolumn{2}{l}{ContFe5}& & & & & &\\ 
& Cont+Fe:& $-1.243 \pm 0.010$ & $(5.22 \pm 0.40)\cdot 10^{-16}$ & 
$2500 \pm 300$ & $0.01195 \pm 0.00053$ & $85$ & 1.71 & ContFe5\\

\hline
7.& \multicolumn{2}{l}{ContFe6}& & & & & &\\ 
& Cont+Fe:&  $-1.215 \pm 0.012$ & $(5.35 \pm 0.43)\cdot 10^{-16}$ & 
$1179 \pm 190$ & $0.01083 \pm 0.00043$ & $74$ & 1.69 & ContFe6\\ 

\hline
\end{tabular}
\\
\begin{flushleft}

Column (1) - model number; column (2) - model name: fits no.~1--4 have
continuum modeled as a power law in spectral windows defined by model
``Cont'' in Table~1, independently from the iron emission fitted in
iron spectral windows defined by models Fe1,2,3 in Table~1; fits
no.~5--7 have continuum and iron emission fit simultaneously to windows
defined by models ContFe4,5,6 in Table~1.

Column (3) lists the power law index, $\alpha_\lambda$ (where
$f_\lambda \propto \lambda^{\alpha_\lambda}$). Column~(4) shows the
amplitude of the continuum measured at 3000 \AA. Columns~(5)-(7) list
rest frame full width at half maximum (FWHM) of the Fe\,II template,
scaling factor, and Fe\,II rest frame equivalent width (REW),
respectively. The reduced $\chi^2$ of the fitting procedure is shown
in Column~(8). Column~(9) lists the fitting windows used by us and
defined in Table~\ref{tab:OurWindows}.
\end{flushleft}
\label{tab:continuum}
\end{table*}

\begin{table*}
\caption{Rest frame absorption lines properties of 
SDSS J094533.99+100950.1
}
\begin{tabular}{rcccccc}
\hline 
\hline
Line & $\lambda_{\rm em}$ & $\lambda_{0}$ & REW& FWHM& Shift& z$_{\rm abs}$\\
& [\AA]  & [\AA]  & [\AA] & [km s$^{-1}$]  & [km s$^{-1}$] &\\
\hline

{\bf fit: Cont-Fe1-best}& & & & & &\\
Mg\,II & 2796.357 & $ 2721.30 \pm 0.51 $ & $ -3.79 ^{ +0.34 }_{ -0.31 }$ & $ 343  \pm  82 $ & $ -8268  \pm  57 $ &  1.59016 \\ 
       & 2803.536 & $ 2728.26 \pm 0.58 $ &  & $ 290  \pm  110 $  & $ -8272  \pm  64 $ &  1.59013 \\ 
C\,IV & 1548.189 & $ 1506.39 \pm 0.22 $ & $ -3.35 ^{ +0.42 }_{ -0.47 }$ & $ 455  \pm  63 $  & $ -8318  \pm  44 $ &  1.58975 \\ 
      & 1550.775 & $  1509.24 \pm 0.24 $ &  & $ 269  \pm  69 $  & $ -8251  \pm  49 $ &  1.59031 \\ 
\hline

\end{tabular}
\\
\begin{flushleft}

Column~(1) - name of absorption line, column (2) - the wavelength of
the line seen in laboratory. The absorption lines were fitted with
Gaussians to a dereddened, rest frame, FeII subtracted quasar
spectrum. Column~(3) - wavelength of the minimum intensity of the line
in the fit.  Column (4) - Rest frame Equivalent Width (REW) of the
doublet (computed as a sum of two Gaussians). Column (5) Full Width at
Half Maximum (FWHM) of the absorption line. Column (6) lists the line
blueshift in \kms .
We present measurements for the Cont-Fe1-best model from Table~2 only,
as other models gave similar results.
\end{flushleft}
\label{tab:absorption}
\end{table*}

\begin{table*}
\caption{Rest frame emission lines properties of 
SDSS J094533.99+100950.1. }

\begin{tabular}{lcccc}
\hline
\hline
Line & $\lambda_{0}$  & REW  & FWHM  & z$_{\rm em}$ \\
& [\AA]   & [\AA] & [km s$^{-1}$] & \\
\hline

{\bf fit: Cont-Fe1-best}& & & & \\
Mg\,II $\lambda2800$ & $2800.29 \pm 0.79$ & $15.3^{+1.1}_{-1.1}$ & $5490
\pm 220$ & 1.66187\\ 
C\,III] $\lambda1909$ + Si\,III] $\lambda1892$& $1906.7 \pm 2.3$ & $1.68^{+0.61}_{-0.52}$ & $4800
\pm 4800$ & 1.65845\\ 
C\,IV  $\lambda1549$ & $1544.9 \pm 2.3$ & $1.49^{+0.82}_{-0.64}$ & $4000
\pm 1100$ & 1.65451\\ 
\hline

fit: Cont-Fe1 & & & & \\
Mg II $\lambda2800$ & $2800.60 \pm 0.81$ & $15.5^{+1.1}_{-1.1}$ & $5690
\pm 230$ & 1.66217\\ 
C\,III] $\lambda1909$ + Si\,III] $\lambda1892$& $1905.6 \pm 2.4$ & $1.69^{+0.63}_{-0.53}$ & $5000
\pm 5000$ & 1.65680\\
C\,IV  $\lambda1549$ & $1545.6 \pm 2.2$ & $1.41^{+0.79}_{-0.62}$ & $3800
\pm 1100$ & 1.65572\\ 
\hline

fit: Cont-Fe2 & & & & \\
Mg\,II $\lambda2800$ & $2800.85 \pm 0.83$ & $15.2^{+1.1}_{-1.1}$ & $5750
\pm 240$ & 1.66240\\ 
C\,III] $\lambda1909$ + Si\,III] $\lambda1892$& $1904.6 \pm 2.5$ & $1.69^{+0.64}_{-0.54}$ & $5100
\pm 5100$ & 1.65539\\
C\,IV  $\lambda1549$ & $1545.9 \pm 2.3$ & $1.36^{+0.78}_{-0.60}$ & $3800
\pm 1100$ & 1.65632\\ 
\hline

fit: Cont-Fe3 & & & & \\
Mg\,II $\lambda2800$ & $2801.64 \pm 0.92$ & $15.9^{+1.2}_{-1.2}$ & $6290
\pm 270$ & 1.66316 \\ 
C\,III] $\lambda1909$ + Si\,III] $\lambda1892$& $1903.9 \pm 2.7$ & $1.64^{+0.66}_{-0.55}$ & $5200
\pm 5200$ & 1.65455\\
C\,IV  $\lambda1549$ & $1545.7 \pm 2.5$ & $1.18^{+0.77}_{-0.58}$ & $3600
\pm 1200$ & 1.65597\\ 
\hline

fit: ContFe4 & & & & \\
Mg\,II $\lambda2800$ & $2799.28 \pm 0.91$ & $14.0^{+1.2}_{-1.1}$ & $5670
\pm 250$ & 1.66091\\ 
C\,III] $\lambda1909$ + Si\,III] $\lambda1892$& $1905.6 \pm 2.5$ & $1.37^{+0.59}_{-0.48}$ & $4400
\pm 4400$ & 1.65680\\ 
C\,IV  $\lambda1549$ & $1543.7 \pm 2.6$ & $1.59^{+0.91}_{-0.71}$ & $4300
\pm 1300$ & 1.65252\\ 
\hline

fit: ContFe5 & & & & \\
Mg\,II $\lambda2800$ & $2804.77 \pm 0.76$ & $16.0^{+1.1}_{-1.1}$ & $5600
\pm 220$ & 1.66614\\ 
C\,III] $\lambda1909$ + Si\,III] $\lambda1892$& $1902.5 \pm 2.5$ & $1.91^{+0.68}_{-0.58}$ & $5400
\pm 5400$ & 1.65252\\
C\,IV  $\lambda1549$ & $1547.4 \pm 2.4$ & $1.11^{+0.73}_{-0.55}$ & $3400
\pm 1100$ & 1.65888\\ 
\hline

fit: ContFe6 & & & & \\
Mg\,II $\lambda2800$ & $2798.84 \pm 0.87$ & $14.1^{+1.2}_{-1.1}$ & $5390
\pm 240$ & 1.66050\\ 
C\,III] $\lambda1909$ + Si\,III] $\lambda1892$& $1906.0 \pm 2.4$ & $1.33^{+0.58}_{-0.47}$ & $4300
\pm 4300$ & 1.65738\\ 
C\,IV  $\lambda1549$ & $1543.7 \pm 2.6$ & $1.45^{+0.89}_{-0.68}$ & $4200
\pm 1300$ & 1.65244\\ 
\hline

\end{tabular}
\\
\begin{flushleft}

Column~(1) - name of emission line, column~(2) - wavelength of the
maximum intensity of the line in the fit (emission lines were fitted
with a Gaussian to a dereddened, rest frame, FeII subtracted quasar
spectrum), column (3) - Rest frame Equivalent Width (REW) of the line,
column (4) - Full Width at Half Maximum (FWHM) of the emission line,
Column (5) redshift. \\


The fitting window for Mg II is 2750--2840\AA, for C\,IV 1530--1580\AA
, and for C\,III] + Si\,III] 1880--1940\AA.
\end{flushleft}
\label{tab:emission}
\end{table*}

\begin{table*}
\begin{center}
\caption{Emission line properties in other quasar
samples.
}
\begin{tabular}{crrrrr}
\hline
\hline
Line  & REW(composite) & REW(FOS) & FWHM(FOS) & REW(LBQS) & FWHM(LBQS) \\
      & [\AA]        &  [\AA]  &  km s$^{-1}$ & [\AA]   &   km s$^{-1}$ \\
\hline
Mg\,{\scriptsize II}  \\ 
single & 30.14 & $136 \pm 92^{1}$ & $3840 \pm 1850$ & $39\pm22$ & $5160 \pm 120$\\
broad  &       & $33  \pm 8$  & $8230 \pm 1850$ & $37\pm22$ & $8660 \pm 300$\\
narrow &       & $29  \pm 9$  & $3260 \pm 760$  & $28\pm13$ & $3510 \pm 110$\\
\hline
C\,{\scriptsize III]} &
21.45  & $21 \pm 5$& $4890 \pm 780$ & $28\pm15$ & $7820 \pm 170$ \\
\hline
C\,{\scriptsize IV}  \\ 
single & 26.86 & $21 \pm 16^{1}$ & $4430 \pm 3070$ & $38\pm20$ & $7720 \pm 150$\\
broad  &       & $63 \pm 7$  & $11090\pm 1070$ & $45\pm 23$& $10960\pm 360$\\
narrow &       & $30 \pm 4$  & $2900 \pm 290$  & $17.8\pm9.3$&$2860 \pm 110$\\
\hline
\end{tabular}
\\
\begin{flushleft}

Emission line parameters of SDSS~J094533.9+100950 are compared with:
in column (2) the Richards et al. (2003) composite no.~3
(Fig.~\ref{fig:spec}) built on the basis of 770 quasar spectra, in
columns (3),(4) the 158 AGN observed with Faint Object Spectrograph on
the Hubble Space Telescopes (FOS; heterogeneous sample of mixed type
AGN, Kuraszkiewicz et al. 2002) and in columns (5),(6) the 993 quasars
from the Large Bright Quasar Survey (LBQS; Forster et al. 2001).  In
Column (1) ``single'' means that the line was fitted by a single
Gaussian, while ``broad'' and ``narrow'' refer to the broad and narrow
components of the line. \\ 
$^1$ - The single Gaussian fits to Mg\,II
and C\,IV emission lines in FOS were based only on a few, low quality
spectra and thus may not be representative.
\end{flushleft}
\label{tab:porownanie}
\end{center}
\end{table*}

The emission and absorption lines were fitted using Gaussians and
the fits were performed for all continuum and iron models defined in
Table~\ref{tab:continuum}.
The different models had no significant effect on the absorption line
fits (resulting in less than 11\% difference in equivalent widths) and
so in Table~\ref{tab:absorption} we give the values for one
model only Cont-Fe1-best.

The Mg\,II emission line was more influenced by the choice of
continuum and iron models, and the equivalent width varied from fit to
fit by a few percent, with a 15\% difference between the Cont-Fe1-best
and the extreme value (see Table~\ref{tab:emission}). The measured
kinematic width depended on the spectral decomposition, and varied
from 5390~\kms up to 6290~\kms, depending on the model used. However,
in all studied cases the line was relatively strong and
broad. Therefore, we consider that the properties of the Mg\,II line
are determined reliably which is important from the point of view of
black hole mass determination.  The C\,IV line is consistently faint
in all models, although some fits indicate the presence of a very weak
line while others give values consistent with zero intensity.

For comparison, in Table~\ref{tab:porownanie} we give the typical
emission line parameters measured for QSOs. Single component fits and
two (narrow and broad) component fits are given whenever possible. We
see that the Mg\,II line intensity is a factor of $\sim 2$ lower than
the mean Mg\,II line intensity of sources found in the Large Bright
Quasar Survey (LBQS, Forster et al. 2001) or observed with the Faint
Object Spectrograph on the Hubble Space Telescopes (FOS, Kuraszkiewicz
et al. 2002) as well as the value found for the single component
Mg\,II fits to the Richards et al. (2003) steep composite.  However,
C\,IV line (as well as C\,III]) is over 20 times fainter. The weakness
of C\,IV - typically one of the strongest quasar lines - qualifies the
object as a WLQ. However, the presence of Mg\,II allows for an
estimate of the global physical parameters of our source.

We also tested the possibility that the redshift determination based
on the Mg\,II line is inaccurate and we allowed the line centroid and
the position of the iron line complexes to shift arbitrarily.  This
did not improve the fit significantly although a possibility of such a
shift by a few hundred \kms cannot be rejected.
	
Since WLQs have been occasionally suggested to be a BAL (McDowell et
al. 1995), we also calculated the BI index (Weymann et al. 1991) in
the vicinity of the Mg\,II line. We obtained a value consistent with
zero. Also the AI index (Hall et al. 2002) for this quasar is zero if
we take the unbinned spectrum where NAL doublets are clearly resolved
- Fig.~\ref{fig:lines}. There is also no sign of a C\,IV BAL, where
the observed spectrum allows detection of such a component blueshifted
up to $\sim$ 20\,000~km~s$^{-1}$.  We conclude that SDSS
J094533.99+100950.1 is not a BAL or a mini-BAL object.

\subsection{Spectral energy distribution}

We compiled the broad band spectral energy distribution (SED) of
SDSS~J094533.99+100950.1 
using the data archives and software of the Virtual Observatory (VO)
\footnote{\texttt{http://www.euro-vo.org/}}
and the Galaxy Evolution Explorer (GALEX) 
\footnote{\texttt{http://galex.stsci.edu/}} database.  This included
optical u, g, r, i and z photometry from SDSS, and near-IR J, H and
$\rm K_{s}$ photometry from the Two Micron All Sky Survey (2MASS) and
near- (NUV) and far-ultraviolet (FUV) photometry from GALEX (Edge,
private communication).

No other spectra or photometric data points were found in the VO
within the matching radius of 5''. All available images of the quasar
and its vicinity were downloaded. The quasar is visible in the SDSS
and 2MASS images, and also on the Palomar Observatory Sky Survey
(POSS) plates in the optical and near-IR. No X-ray data (XMM-Newton,
RXTE, INTEGRAL) were found.  We searched for X-ray spectra in the
ROSAT catalogues, but found no sources with a flux significantly
higher than the X-ray background (So\l tan, private
communication). The radio properties of our quasar and its
neighbourhood were checked in the Parkes-MIT-NRAO (4850 MHz survey),
and in the Faint Images of the Radio Sky at Twenty-centimeters (VLA
FIRST, 1.4 GHz).  In both cases the radio background was found and the
spatial resolution was too low to observe the quasar. The same results
were obtained for IRAS and EUVE catalogues - the signal did not differ
significantly from the background.

\begin{figure}
\epsfxsize=8cm
\epsfbox{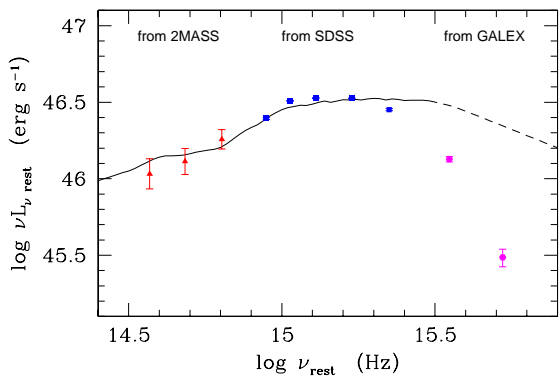}
\caption{
The broad band SED of SDSS J094533.99+100950.1. The photometric IR
data point (triangles) come from 2MASS (J,H,K$_{\rm s}$ colours).
Optical photometry (squares) is from SDSS (u,g,r,i,z filters). The NUV
and FUV points (hexagons) are from GALEX.  The spatial coordinates of
the source differ between SDSS and 2MASS by $\sim 0.39$ arcsec and
between SDSS and GALEX by $\sim 0.48$ arcsec. We assume the data from
SDSS, 2MASS and GALEX come from the same object. Galactic dereddening
was done using the values E(B-V)=0.060, R$_{\rm V}$ = 3.2 and formulae
of Cardelli et al. (1989).  Neither intrinsic dereddening of the
quasar nor subtraction of Fe\,II were done.  For comparison the line
shows the SED from Elvis et al. (1994) marking with dashed line the
frequency range of interpolation between the far UV and soft X-ray
band.
}
\label{fig:totspec}
\end{figure}

The SED of SDSS~J094533.99+100950.1 is shown in
Fig.~\ref{fig:totspec}.  It does not differ significantly from typical
quasar SEDs (e.g. Elvis et al. 1994, Richards et al. 2006), consistent
with the conclusion reached by Diamond-Stanic et al.  (2009) for other
WLQs. The deviation from the Elvis et al. (1994) median SED in the far
UV (at NUV and FUV GALEX data points) is due to the simple
interpolation between the UV and soft X-rays in the Elvis et
al. (1994) SED, which does not constrain the true QSO SED at
$\lambda<1000$~\AA.  
However, lower flux in the far-UV remains a
possibility in our WLQ, which may lead to low CIV emission due to
lower ionizing flux.

\section{Global parameters of SDSS J094533.99+100950.1}
\label{sect:global}

Our WLQ has a monochromatic luminosity at 3000\AA\ of $\lambda
L_{\lambda} (3000 \rm \AA) = 2.95 \times 10^{46}$ erg s$^{-1}$,
(assuming $H_0 = 71 $ km s$^{-1}$ Mpc$^{-1}$, $\Omega_{\Lambda} =
0.73$ and $\Omega_{m} = 0.27$ - Spergel et al. 2007), which is high
but not extremely high for a quasar.
Using a bolometric correction BC$_{\rm 3000} = 5.15$ determined at
3000\,\AA\ (Shen et al. 2008 and Labita et al. 2009, based on Richards
et al. 2006) we obtain a bolometric luminosity of $1.52 \times
10^{47}$ erg s$^{-1}$.

We use the following equations from Kong et al. (2006):
to determine the central black hole mass: 

\begin{equation}
M_{\rm BH} =  3.4 \times 10^6 \Big(\frac{\lambda L_{\rm 3000
\AA}}{10^{44} \rm erg s^{-1}}\Big)^{0.58\pm0.10} 
\Big[\frac{\rm FWHM_{\rm MgII}}{1000 \, \rm km s^{-1}} \Big]^2
M_{\odot}
\label{eq:Mbh3000}
\end{equation}
\begin{equation}
M_{\rm BH} =  2.9 \times 10^6 \Big(\frac{L_{\rm MgII}}{10^{42} 
\rm erg s^{-1}}\Big)^{0.57\pm0.12} 
\Big[\frac{\rm FWHM_{\rm MgII}}{1000 \, \rm km s^{-1}} \Big]^2
M_{\odot}
\label{eq:MbhMgII}
\end{equation}
where $\lambda L_{\rm 3000 \AA}$ and $L_{\rm Mg II}$ are the
monochromatic continuum luminosity at 3000~\AA\ and the Mg\,II line
intensity, respectively.  Using the Mg\,II width from the best fit
(Cont-Fe1-best in Table~\ref{tab:emission}) we obtain a black hole
mass of $2.7 \times 10^9 M_{\odot}$ from
equation~(\ref{eq:Mbh3000}). This is a typical value for a distant
quasar (e.g. Kelly et al. 2008, Vestergaard et al. 2008).  Using
equation~(\ref{eq:MbhMgII}) we obtain a slightly lower value of the
black hole mass, of $1.5 \times 10^{9} M_{\odot}$. Similar values of
M$_{\rm BH}$ are obtained when using formulae from McLure \& Dunlop
(2004).

The first value of the black hole mass implies the Eddington ratio of
0.45. This value is very conservative from the point of view of
possible super-Eddington accretion, which is possibly present in some
AGN (Collin \& Kawaguchi 2004, Jin et al. 2009). It is consistent with
several other determinations of this ratio in quasars. Quasars in the
redshift range between 1 and 2 in the study of Sulentic et al. (2006)
have Eddington ratios between $\sim$ 0.2 and $\sim$ 1, the value 0.25
is the mean favoured by statistical studies of Shankar et al. (2008).
Labita et al. (2009) found a value of 0.45 as a statistical upper
limit of accretion rate in their sample of quasars.  The second value
based on black hole mass from equation~(\ref{eq:MbhMgII}) gives a
somewhat higher ratio of 0.79, which is less reliable because of the
weakness of the Mg\,II emission line.  We also computed the mean and
standard deviation of the Eddington ratio for all models in
Table~\ref{tab:continuum} and obtained the following means: $0.429 \pm
0.037$ using equation~(\ref{eq:Mbh3000}) and $0.752 \pm 0.077$ for
equation~(\ref{eq:MbhMgII}).  In the newest studies of the relation
between Mg\,II line parameters and black hole mass - Wang et
al. (2009) do not use Mg\,II equivalent width but rather the Mg\,II
FWHM and continuum luminosity at 3000\AA. 
Using the relation from Wang et al. (2009) the mean black hole mass of
SDSS J094533.99+100950.1 is $(3.09 \pm 0.23) \times 10^{9} M_{\odot}$
and the mean Eddington ratio is $0.396 \pm 0.026$.

Additionally, since the Mg\,II emission line intensity is lower than
in typical quasars, the second method is probably less reliable than
the first, based on the continuum, and so the Eddington ratio of 0.45
is more likely.

\section{Discussion}

The weak line quasar SDSS J094533.99+100950.1 is an exceptional object
with almost nonexistent C\,IV emission, normal Fe\,II emission and
visible, although somewhat fainter and atypical, Mg\,II emission
line. The presence of iron emission with standard intensity argues
against any relativistic enhancement of the continuum in this
source. The quasar continuum is normal, similar to the steep composite
no.~3 of Richards et al. (2003).

The lack of strong emission lines (except for Mg\,II) is likely to be
intrinsic, consistent with the conclusion reached by Diamond-Stanic et
al. (2009) for other WLQs. A remarkable property of the SDSS
J094533.99+100950.1 spectrum is the presence of Low Ionization Lines
(LILs) such as Mg\,II, which are thought to form close to the
accretion disk surface (see e.g. Collin-Souffrin et al. 1988) and the
absence of High Ionization Lines (HILs) such as C\,IV. The narrow
component of Mg\,II, which is thought to form at larger distances from
the nucleus, is also absent from the spectrum.

McDowell et al. (1995) listed 10 possible explanations for the
weakness of emission lines in their WLQ, PG~1407+265, but rejected
most of them on the basis of overall quasar properties. Among those
remaining are: high extinction or BAL effects, which are inconsistent
with SDSS J094533.99+100950.1 properties, and abnormal ionizing
continuum, which remains a possibility. In particular, the option of
high (super-Eddington) accretion rate may seem attractive 
since the determination of the Eddington ratio for our object
allows (but does not require) for such a possibility (see
Sect.~\ref{sect:global}). However,
the expected trends found by PCA analysis performed on the optical/UV
emission line properties of PG QSO (Shang et al. 2003; see Fig.~10)
show that, with increasing Eddington ratio, the UV Fe\,II emission
(measured around Mg\,II) decreases\footnote{not increases, as is the
case for optical Fe\,II measured around H$\beta$}, Si\,III]+C\,III]
emission increases, and the widths of Mg\,II and Si\,III]+C\,III]
lines decrease. The opposite is seen in our object. Additionally the
Si\,IV line should become stronger, but this is not seen in the WLQs
of Shemmer et al. (2009). Since the emission line properties of WLQs
do not follow the trends for high $L/L_{Edd}$ objects predicted by PCA
analysis, something different than high $L/L_{Edd}$ is causing the
atypical emission line behaviour.

The only explanation, which in our opinion is currently consistent
with the intriguing emission line properties of our WLQ is that the
quasar activity in this source has just begun.

The formation of an accretion disk in an AGN is a long time
phenomenon, occurring in millions of years (the viscous timescale in
the outer disk, Siemiginowska et al. 1996). However, when the disk
finally approaches a black hole, i.e. when the inner radius moves from
$\sim 10 - 20 R_{\rm Schw}$ to ISCO (Inner Stable Circular Orbit), the
disk's luminosity rapidly increases on a hundred year timescale and
X-ray emission starts (Czerny 2006).

The immediate stage after is the irradiation of the outer disk which
happens in the light travel time across the region (years). At this
stage the accretion disk continuum appears already to be typical for
an AGN. However, neither the broad-line region (BLR) nor the
narrow-line region (NLR) region, exist.

If our explanation is correct, our WLQ should have no narrow
[O\,III]$\lambda$5007 emission but its LIL H$\beta$ line should have
already formed (as has the Mg\,II line). For our $z = 1.6$ object
these lines lie outside the observed wavelength range of the optical
telescopes used by SDSS, but could, in principle, be observed using an
IR facility. In another WLQ, PG~1407+265, (McDowell et al. 1995) the
H${\beta}$ line is weak, contrary to our expectations.  However,
significant influence of the atmosphere above 9000\AA\ leading to low
spectrum quality at those wavelengths may partially hide the line (see
Fig.~1 of McDowell et al. 1995). The puzzling properties of the X-ray
emission of PG 1407+265 may be consistent with the young age of the
source as the emission significantly varies in time, and is only
occasionally dominated by a jet (Gallo 2006) indicating that a stable
jet has not yet been launched.
 
A relatively young age was also suggested for the class of Narrow Line
Seyfert 1 (NLS1) galaxies, in the context of cosmological evolution
(Mathur 2000) but their BLR is well developed. WLQs thus possibly
represent an even earlier stage but not necessarily in the sense of a
global evolution since the argument of their young age is based on BLR
formation stage. If the observed activity is the first active stage of
the nucleus then indeed the WLQ stage should be seen first, later
replaced by the NLS1 stage (with efficient growth of the central black
hole), followed by the broad emission line stage (quasar or Seyfert
1). However, it is also possible that the observed active stage
represents the source's reactivation. The amount of intrinsic
(nuclear) dust along the line of sight does not seem to be large and
the past activity may be the natural explanation for such a clean
environment for an otherwise young object. The large mass of the black
hole, particularly in PG 1407+265 ($\log M_{\rm BH} = 9.8$ based on
C\,IV FWHM and the formula of Vestergaard \& Peterson 2006) might
rather indicate a relatively late evolutionary stage, favouring the
reactivation scenario.  A few stages of quasar life are also predicted
by Hopkins \& Hernquist (2009). They postulate that the life of
quasars consists of a merger and a few episodes of high accretion
rate.  The activity is likely to cease through a ``naked'' AGN stage
(Williams et al. 1999, Hawkins 2004) since when the accretion rate
drops, the optical disk recedes outward and the component of the BLR
linked to the disk wind disappears (e.g. Czerny et al. 2004, Elitzur
\& Ho 2009).

An intermittent character for the quasar activity is also more
consistent with the statistics of WLQ. The sources are rare but not as
rare as we might expect (based on the following discussion). 
The irradiation causes the formation of a
wind in the upper disk atmosphere, and this material slowly rises up
in the vertical direction. This wind is customarily now considered as
the source of the material from BLR clouds (Murray et al. 1995; Proga
et al. 2000; Chelouche \& Netzer 2003; Everett 2005).  The initial
rise is not highly supersonic. Assuming velocities
$\sim$ 100~km~s$^{-1}$ (e.g. Proga 2003, Everett 2005) we can estimate
how much time the material takes to depart significantly from the
disk's surface layers.  Assuming a disk radius of 1 pc, a rise by ten
percent of the radial distance takes 1000 years, and at that stage the
low ionization lines forming close to the disk, like Mg\,II and
H$\beta$ appear.  The rise to the height comparable to the radial
distance takes a factor 10 times longer, and only after that time the
highly ionized lines such as C\,IV appear in the spectrum.  The narrow
emission line components take longer to form and are likely to be
absent at the initial stages. A similar argument was used for the
explanation of the relative faintness of the [O\,III] line in GPS
quasars which are also thought to be relatively young (Vink et
al. 2006). Thus, a WLQ phase lasts for about 1000 years.  There are
almost 100 quasars classified as WLQ (Shemmer et al. 2009,
Diamond-Stanic et al. 2009, Plotkin et al. 2010) in the whole sample
of $\sim$ 100 000 SDSS quasars. However only a fraction of those,
including SDSS J094533.99+100950.1 (this paper), PG 1407+265 (McDowell
et al. 1995) and HE 0141-3932 (Reimers et al. 2005), may represent a
more advanced stage with partially developed low ionization
lines. Therefore, the probability of observing this evolutionary stage
can be estimated as $10^{-3}-10^{-4}$.
This would indicate that
the typical duration of a quasar's active phase is only $\sim 10^6-10^7$
years.  On the other hand, the lifetime of the luminous phase of
quasars is $\sim 10^7 < t_{\rm QSO} < \sim 10^8$ yrs (Haiman \& Hui,
2001), but in some sources extends over a period $\sim 10^9$ yrs
(Martini \& Weinberg, 2001). The two numbers can be reconciled if the
total active phase of a quasar consists typically of 100 separate
subphases, each of those starting with a WLQ stage of BLR buildup.

The increasing probability of occurrence of WLQ among high redshift
objects (Diamond-Stanic et al. 2009) agrees with our hypothesis if, in
the earlier Universe, minor mergers are more frequent and the typical
single active subphase lasts shorter.

Thus the SDSS J094533.99+100950.1 quasar with its remarkable
properties may have a Rosetta stone impact on our understanding of
quasar activity cycle and the nature of WLQ.

Finally, it is important to note that the definition of a WLQ has not
been yet worked out. Some objects with extremely faint/nonexistent
lines, may indeed be BL Lacs. Others, like PHL 1811 (Leighly et
al. 2007), considered as a WLQ prototype, may be explained by higher
Eddington ratio since the lines are not only fainter but also narrower
(Mg\,II line of $2550\pm 110$~\kms\ vs. $5490\pm 220$~\kms\ in
J094533.99+100950.1). Our hypothesis of young/newly active sources
would not apply to those classes of objects.

\section{Conclusions}

Quasar SDSS J094533.99+100950.1 is an exceptional example of the Weak
Line Quasar class. The broad band spectrum is typical for quasars,
very similar to the second steepest composite spectrum of Richards et
al. (2003). It has a relatively well developed broad Mg\,II line and
the typical Fe\,II contribution but its high ionization lines, like
C\,IV, are weak/absent.  The BAL phenomenon is not visible in the
spectrum so the lack of C\,IV line is an intrinsic property of the
source. The presence of Mg\,II and typical Fe\,II emission excludes a
non-thermal, blazar-like origin of the spectrum.  The significant
width of Mg\,II line (FWHM of 5500 \kms) and the Fe\,II properties
argue in turn against an extremely high accretion rate in this object.
Objects like these are difficult to find since a typical search for
WLQs, based on finding spectra with extremely weak or non-existent
emission lines is likely to miss sources where one of the lines is
relatively strong. However, such objects are important to find/study
as they are likely to shed light on the the mechanism driving all (or
at least some) WLQs.

In the present paper we propose the following hypothesis for the
nature of WLQ.  AGN are generally classified according to their
viewing angle and Eddington ratio. However, those objects evolve and
some evolutionary stages may be too short to produce a consistent
equilibrium between various emitting components in the nucleus like
the BLR, NLR and the accretion disk. It is an interesting possibility
that the WLQ class represents such a stage, at the onset of quasar
activity. The relatively high occurrence of these sources may
additionally suggest that a quasar's active phase has an intermittent
character. The whole active phase consists of several subphases, each
starting with a slow development of the BLR region which manifests
itself as a WLQ phenomenon. Further observational tests of this
hypothesis are clearly needed.

\section*{Acknowledgments}
We would like to thank an anonymous referee for useful comments that 
improved our paper.
We are grateful to Marianne Vestergaard for providing us with the iron
emission template. We would also like to thank Andrzej So\l tan for
data search in ROSAT catalogues, Alastair Edge for helpful remarks
about GALEX and 2MASS missions and Belinda Wilkes for advices during
improving this paper. 
This work makes use of EURO-VO
software, tools or services. The EURO-VO has been funded by the
European Commission through contract numbers RI031675 (DCA) and 011892
(VO-TECH) under the 6th Framework Programme and contract number 212104
(AIDA) under the 7th Framework Programme.  This work was supported in
part by N N203 380136 and by the Polish Astroparticle Network
621/E-78/BWSN-0068/2008.

\ \\
This paper has been processed by the authors using the Blackwell
Scientific Publications \LaTeX\ style file.

\end{document}